\documentclass[prl,superscriptaddress,twocolumn]{revtex4-1}

\usepackage{graphicx,amsmath,amssymb,amsfonts}   
\usepackage{dcolumn}    
\usepackage{bm}         
\usepackage{color}
\usepackage{times}
\usepackage[bold]{hhtensor}

\newcommand{\etal}{\emph{et al.\@}}

\begin{document}
\author{M. R. Gilbert}
\affiliation{EURATOM/CCFE, Culham Science Centre, Abingdon, Oxfordshire, OX14 3DB, UK}
\author{P. Schuck}
\affiliation{Oak Ridge National Laboratory, Oak Ridge, TN 37831}
\author{B. Sadigh}
\affiliation{Lawrence Livermore National Laboratory, Livermore, CA 94551}
\author{J. Marian}\email{marian1@llnl.gov}
\thanks{Corresponding author}
\affiliation{Lawrence Livermore National Laboratory, Livermore, CA 94551}

\title{Free energy generalization of the Peierls potential in iron}

\begin{abstract}
In body-centered cubic (bcc) crystals, $\frac{1}{2}\langle111\rangle$ screw dislocations exhibit high intrinsic lattice friction as a consequence of their non-planar core structure, which results in a periodic energy landscape known as the Peierls potential, $U_P$.
The main features determining plastic flow, including its stress and temperature dependences, can be derived directly from this potential, hence its importance.
In this Letter, we use thermodynamic integration to provide a full thermodynamic extension of $U_P$ for bcc Fe. We compute the Peierls free energy path as a function of stress and temperature and show that the critical stress vanishes at 700K, supplying the qualitative elements that explain plastic behavior in the athermal limit.
\end{abstract}
\maketitle

Dislocations are ubiquitous line defects that mediate plastic deformation in crystalline materials.
In body-centered-cubic (bcc) metals, plasticity is governed by the motion of $\frac{1}{2}\langle111\rangle$ screw dislocations on close-packed planes. Generally, this motion is understood to occur over a periodic energy landscape known as the \emph{Peierls} potential $U_P$. Theoretical descriptions of this potential show that it is very stiff in bcc Fe, leading in some cases to critical stresses (those at which the lattice resistance is suppressed) in excess of one GPa. However, experimentally it is found that the flow stress ---the macroscopic equivalent of the critical stress--- is roughly one third lower than calculated values. The most convincing explanation for this discrepancy that we possess currently is the contribution of zero-point motion to the Peierls potential at temperatures where quantum effects cannot be neglected \cite{proville2012}.

\begin{figure}[h]
\center
{\includegraphics[width=0.9\linewidth,trim=0cm .30cm 0cm 0cm,clip=true]{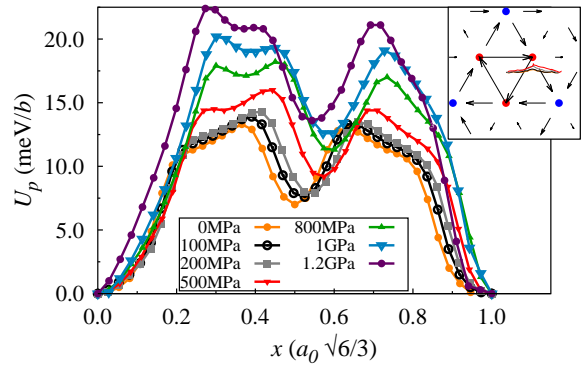}}
\vskip-0.5cm\caption{\label{NEB_stressed}Peierls potential $U_P$  as a function of stress for the Mendelev interatomic potential for Fe \cite{mendelevhan2003}. $x$ represents the (non dimensional) dislocation core position. Note that DFT calculations predict a sinusoidal profile with an amplitude of $\Delta U_P=30\sim40$ meV/$b$ \cite{ventelon2013}. The inset to the figure shows the transition path on the (111) plane taken by the dislocation at several values of $\sigma$ shown in the main figure.}
\label{fig:up}
\end{figure}

At low stresses, one can safely assume that the Peierls potential remains unchanged and that slip proceeds via the thermally-activated nucleation of \emph{wiggles} on the dislocation line, known as kink pairs, and their subsequent sideward relaxation. However, at stresses approaching the critical stress, referred to as \emph{Peierls stress} $\sigma_P$ at 0K, it is insufficient to consider only the zero stress internal energy to represent the Peierls trajectory. This trajectory is defined as the rectilinear path, denoted by the reaction coordinate $x$, between two equivalent equilibrium states (known as `easy core') on the Peierls potential, which has periodicity $h=a_0\sqrt{6}/3$, where $a_0$ is the lattice constant. Rodney and Proville \cite{proville2009} showed that, at moderate to high stresses, the core undergoes internal transformations that modify the Peierls energy landscape. This $U_P(\sigma)$, where $\sigma$ is the shear stress resolved on a \{110\} plane --applied using the Parrinello-Rahman method \cite{parinellorahman1981}--, is provided in Fig.\ \ref{NEB_stressed} using the nudged-elastic-band (NEB) method~\cite{jonssonetal1998} and a standard semi-empirical interatomic force field developed by Mendelev {\it et al.}~\cite{mendelevhan2003}. Although this force field yields the non-degenerate screw dislocation core structure, in accordance with density functional theory (DFT) calculations, it also predicts a metastable \emph{split} core configuration known now to be an artifact~\cite{ventelon2007}. The Peierls potential is then defined by the system enthalpy $H_P(\sigma)=U_P(\sigma)-W_p$, where $W_p$ is equal to the plastic work associated with the shear strain in the simulation box. For an accurate calculation of $W_p$, the reaction coordinate must be expressed in terms of the dislocation core position, which is calculated here by matching the atomic displacement field  to the Volterra solution for a screw dislocation. For the model used in Fig.\ \ref{NEB_stressed}, $\sigma_P$ is approximately 1250 MPa as first computed by Chaussidon {\it et al.}~\cite{chaussidonetal2006} (DFT calculations: $\approx1100$ to $1400$ MPa \cite{mitsu,ventelon2013}).
The inset to Fig.\ \ref{NEB_stressed} shows the two-dimensional representation of the NEB trajectory at several stresses on the $(111)$ plane using a differential displacement map. The figure shows that, as the shear stress increases, the path approaches one of the $\langle111\rangle$ atomic rows associated with split core configuration. At zero stress, the transition path is practically rectilinear --consistent with recent DFT calculations \cite{mitsu,ventelon2013}-- which reflects the structure of the energy landscape for the Mendelev potential \cite{mark}.

Tensile tests place the \emph{athermal} limit of bcc Fe,  {\it i.e.} the point at which flow occurs without mechanical aid, at various temperatures between 300 and 400K \cite{basinski1960,mordike1962,conrad1962,spitzig1970,kuramoto1979,brunner1991}. This limit is thought to establish the extent of validity of the classical kink-pair mechanism. Kink-pair energies have been calculated as a function of stress utilizing atomistic and line tension models, all of which make use of a \emph{substrate} Peierls potential \cite{mitsu,proville2013}.
However, despite its importance, the effect of temperature on the Peierls potential has not yet been addressed.
In this Letter we generalize the Peierls enthalpy to finite temperature conditions by calculating the Gibbs free energy of atomistic Fe systems using a combination of periodic (dipole) and cylindrical configurations containing in excess of $N=12000$ atoms, as described in Refs.~\cite{schuck} and \cite{rao1998}. Dislocation segments $5b$ in length were considered, where $b=a_0\sqrt{3}/2$ is the Burgers vector's modulus and $a_0\approx0.27$ nm.
The reference configurations used here are those calculated at constant stress using the NEB method.

\begin{figure}[h]
\center
{\includegraphics[width=0.9\linewidth,trim=0cm 0.3cm 0cm 0cm,clip=true]
{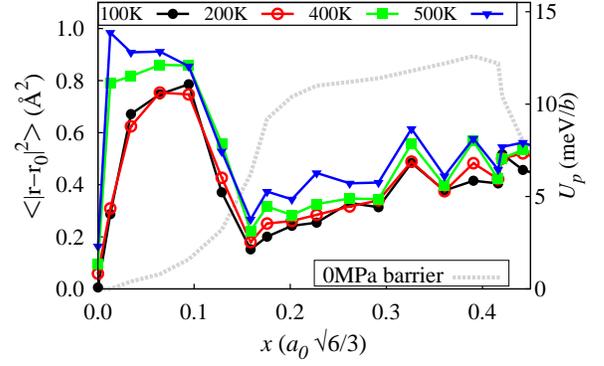}}
\vskip-0.5cm\caption{\label{fig:anh}Time-averaged atomic mean square displacement along the Peierls trajectory (up to 45\% of the reaction coordinate) at zero stress. Areas with large $\langle r^2\rangle$ indicate anharmonic behavior. The unstressed \(U_P\) is shown as a gray dashed line for reference.}
\end{figure}

Adding $T$ to the natural variables $\sigma$ and $N$, results in the isothermal/isobaric ensemble, whose characteristic state function is the \emph{Gibbs} free energy:
$$G=H-TS,$$
where $S$ is the entropy, defined in our pure and periodic systems solely by vibrational contributions. Our objective is to establish the importance of incorporating temperature effects into models based on the standard picture of the Peierls potential and to compute explicity the athermal limit from atomistic calculations.
To obtain the Peierls free energy $G_P$ we equate the Peierls transition path to an activated process described by a general configurational nonlinear many-body reaction coordinate \cite{carter1989}. As a first approximation, we first calculate the harmonic free energy of each configuration along the NEB trajectory as:
\begin{equation}
\label{harmfreeenergy}
G^{h}_i(T)=kT\sum_{\vec{k}}2\sinh{\left(\frac{\hbar\omega_i(\vec{k})}{2kT}\right)}
\end{equation}
where $\omega(\vec{k})$ are the eigenfrequencies corresponding to eigenvectors $\vec{k}$ pertaining to each NEB replica $i$. However, as Fig.\ \ref{fig:anh} demonstrates, relatively large mean square displacements can be measured already at 100K along the reaction coordinate, particularly within the initial 15\% of the Peierls trajectory. Such anharmonic behavior is a manifestation of a pathology of the Mendelev force field, which results in a transition path at 0K that may not reflect the true finite temperature dynamics of the system. 
$G^{h}$ has been calculated for comparison nonetheless, and, following Proville \etal~\cite{proville2012}, to account for zero-point motion at low temperatures. We shall discuss this correction in later paragraphs.

Our method to compute full, anharmonic free energies is based on Kirkwood's approximation to obtain the potential of mean force \cite{kirkwood1935}. Assuming a Hamiltonian of the type ${\cal H}=\frac{\mbox{\boldmath$p$}^2}{2m}+H(\mbox{\boldmath$q$})$, where $\mbox{\boldmath$p$}$ and $\mbox{\boldmath$q$}$ are the generalized momenta and coordinates, respectively, one can write the Gibbs free energy as:
\begin{equation}
\label{free1}
\begin{split}
G&=-\frac{\log{\cal Z}}{\beta}=-\frac{1}{\beta}\log\int{d\mbox{\boldmath$q$}d\mbox{\boldmath$p$}\exp{\{-\beta{\cal H}(\mbox{\boldmath$q$},\mbox{\boldmath$p$})\}}}\\
&=-\frac{{\cal C}}{\beta}\log\int{d\mbox{\boldmath$q$}\exp{\{-\beta U(\mbox{\boldmath$q$})\}}},
\end{split}
\end{equation}
where ${\cal Z}$ is the canonical partition function, $\beta=1/kT$ ($k$ is Boltzmann's constant), and ${\cal C}$ is a constant that represents the integrated contribution of the kinetic energy and the elastic work. If one now extracts the (NEB) trajectory degree of freedom (DOF) $x$ from the $3N$-dimensional vector $\mbox{\boldmath$q$}$, and separate them in eq.\ \ref{free1} we have:
\begin{equation}
\label{free2}
G=-\frac{{\cal C}}{\beta}\int{dx}\left[\log\int{d\mbox{\boldmath$q$}'\exp\{-\beta U(\mbox{\boldmath$q$}',x)\}}\right]
=\int{dx~{\cal G}(x)},
\end{equation}
where $U'=U(\mbox{\boldmath$q$}',x)$ and ${\cal G}(x)$ are the internal and free energies for the $(3N-1)$-DOF system defined by generalized coordinates $\mbox{\boldmath$q$}'$.
The force along the trajectory can be evaluated as:
\begin{equation}
\label{free3}
-\frac{d{\cal G}}{dx}= \frac{{\cal C}}{\beta}\frac{d}{dx}\left[\log\int d\mbox{\boldmath$q$}'\exp\{-\beta U(\mbox{\boldmath$q$}',x)\}\right]
\end{equation}
The quantity ${\cal S}(x)={\cal C}\int d\mbox{\boldmath$q$}'\exp\{-\beta U(\mbox{\boldmath$q$}',x)\}$ can be regarded as the configurational partition function of the $(3N-1)$-DOF system, and, therefore, eq.\ \ref{free3} can be written as:
\begin{equation}
\label{free4}
-\frac{d{\cal G}}{dx}= \frac{1}{\beta}\frac{d\log{\cal S}(x)}{dx}=\frac{1}{\beta{\cal S}(x)}\frac{d{\cal S}(x)}{dx}
\end{equation}
Furthermore,
$$
\frac{d{\cal S}(x)}{dx}=-\beta{\cal C}\int d\mbox{\boldmath$q$}'\exp\{-\beta U(\mbox{\boldmath$q$}',x)\}\frac{dU(\mbox{\boldmath$q$}',x)}{dx},
$$
which, when inserted into eq.\ \ref{free4}, results in:
\begin{equation}
\label{free5}
\frac{d{\cal G}}{dx}=\frac{\int d\mbox{\boldmath$q$}'\exp\{-\beta U'\}\frac{dU'}{dx}}{\int d\mbox{\boldmath$q$}'\exp\{-\beta U'\}}=\left\langle\frac{dU'}{dx}\right\rangle,
\end{equation}
which is a configurational average over all $(3N-1)$ DOF. In other words, the free energy of the constrained system is obtained by integrating the time-averaged total force along the minimum free energy path.

\begin{figure}[h]
\center
{\includegraphics[width=1.05\linewidth,trim=0cm 0.3cm 0cm 0cm,clip=true]{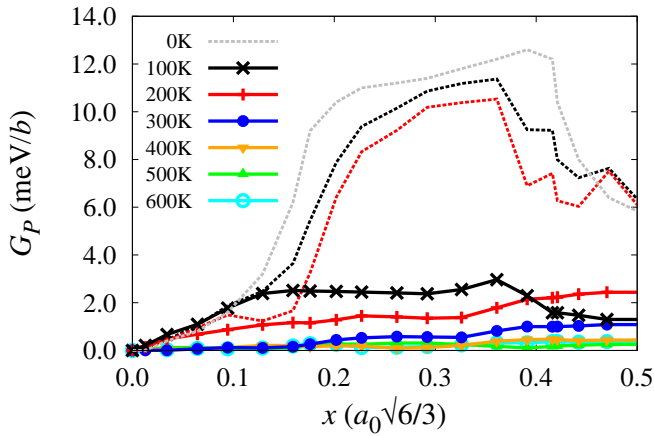}}
\vskip-0.5cm\caption{\label{free_energy_barriers_0MPa}Peierls free energy path at zero stress as a function of temperature. The harmonic free energies at 100 and 200K are also shown as dashed lines for comparison. At 700K and above the barrier is at or below zero.}
\end{figure}

Using eq.\ \ref{free5}, we now calculate free energies for the different trajectory points of the NEB calculations in our atomistic systems. Configurational averages are numerically intensive, and require long simulation times to converge (on the order of several ns in our case). The resulting free energies for the unstressed configurations are shown in Fig.~\ref{free_energy_barriers_0MPa} at temperatures ranging from 100 to 600K. $G^{h}_P$ at 0, 100 and 200K are also included for comparison. Two features are noteworthy at first glance: (i) the characteristic metastability associated with the split core configuration at 0K is lost following constrained equilibration at finite temperatures; (ii) the free energies suffer a marked decrease from 0 to 100K. We find that $\Delta G_P$ vanishes completely by 700K at zero stress. Technically, the current force field for Fe is strictly valid above the Debye temperature and below the Curie transition ($470\lesssim T\lesssim1040$K).
However, Proville \etal~have shown that quantum effects in this context are only important below 40K \cite{proville2012}, and so our results over the 100-to-700K temperature range are within the validity margins of the potential.

As mentioned above, the double-`hump' shape of the Mendelev interatomic force field as given by the NEB method is a consequence of the metastability of the split core configuration at 0K. By contrast, DFT calculations show that the energy path displays a simpler sinusoidal profile \cite{ventelon2007,mitsu}. Gilbert \etal~have performed a detailed numerical construction of the two-dimensional energy landscape for the Mendelev potential on the \{111\} plane \cite{mark}. Their analysis revealed no justification for the standard NEB trajectory to be favored at 0K. Thus, at finite temperatures the system escapes the NEB path and samples a broader region of phase space until falling into an alternate dynamic path. This path is the one shown in Figure \ref{free_energy_barriers_0MPa} at different temperatures.
This behavior explains the long convergence times and the large anharmonicities captured in Fig.\ \ref{fig:anh} as the system samples alternative phase space trajectories. In any case, at 100K both the harmonic and full free energy curves mimic one another for 0$<$$x$$<$0.15, which indicates that within that interval the minimum potential and free energy paths are equivalent. At higher temperatures, the range of agreement gradually decreases, as more and more of the barrier is subject to anharmonic effects.

\begin{figure}[h]
\center
{\includegraphics[width=1\linewidth,angle=0,trim=0cm 0cm 0cm 0cm,clip=true]{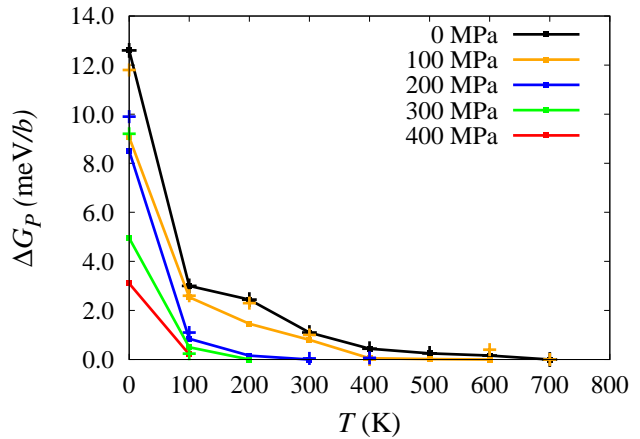}}
\vskip-0.3cm\caption{\label{DeltaG}Variation of the free energy barrier $\Delta G_P$ with temperature and stress. Solid lines correspond to $\Delta G_P(0)$, which is valid up to $\approx$400 MPa. Scatter points represent $\Delta G_P(\sigma)$ --{\it i.e. } the free energies for the stress dependent potential obtained from NEB. In the latter case, the free energy is strictly zero for $\sigma\ge400$ MPa.}
\end{figure}

Next, we study the variation of the free energy barrier $\Delta G_P$ with temperature and stress.
When $U_P(\sigma)$ displays a weak or no dependence on the stress, the dependence of Peierls enthalpy on $\sigma$ is via the temperature-independent plastic work $W_p=\sigma bhx$ (per unit length).
This is shown to be an accurate approximation for $\sigma<500$ MPa (cf.\ Fig.\ \ref{NEB_stressed}). In such case one need only consider the temperature dependence of the unstressed Peierls trajectory (given in\ Fig.\ \ref{free_energy_barriers_0MPa}) and subtract $W_p$ to obtain $G_P(\sigma;T)$. From this, $\Delta G_P$ is measured at each stress and temperature and each value plotted in Fig.\ \ref{DeltaG}. We term this approximation $G_P(0)$ (or $\Delta G_P(0)$ if referring to free energy barriers). The figure reveals several interesting features. First, the free energy decreases by more than 50 percent from zero to 100K. Subsequently, it decreases gradually until it vanishes. This latter point furnishes the critical stress $\sigma^{*}$, {\it i.e.} that at which $\Delta G_P(\sigma^{*};T)=0$. Second, the curves at different stresses roughly mimic one another within the envelope of the zero stress results, which is an indication that the dynamic path is stable. Much in the manner of Proville {\it et al.} \cite{proville2012}, adding zero-point corrections reduces the value of $\Delta G_P$ at low temperatures for all stresses. For clarity, this effect is not shown in the figure but will be taken into account when computing the critical stresses.

\begin{figure}[h]
\center
{\includegraphics[width=1.0\linewidth,trim=0cm 0cm 0cm 0cm,clip=true]{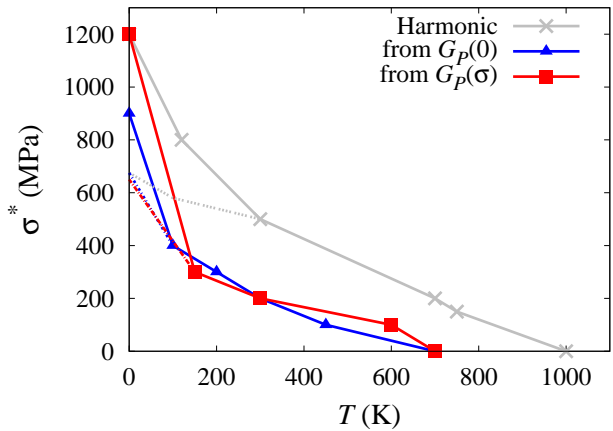}}
\vskip-0.3cm\caption{\label{critical}Variation of the critical stress $\sigma^{*}$ with temperature from full and harmonic free energy calculations. Solid lines represent classical calculations, whereas the dashed ones symbolize the quantum corrections at low temperature. In the classical limit, $\sigma_P\equiv\sigma^*$ at $T=0$K.}
\end{figure}

To verify the assumption of stress-independence for the Peierls potential at 0K below $\approx500$ MPa, we have  performed free energy calculations on stressed transition paths at stresses 100 to 400 MPa. The results as a function of temperature are shown as scatter points in Fig.\ \ref{DeltaG}. As shown, the agreement between $\Delta G_P(0)$ and the stressed configurations ($\Delta G(\sigma)$) is reasonable up to 300 MPa. From the data points presented in Fig.\ \ref{DeltaG}, we can extract the values of $\sigma^{*}$ at each temperature. These are shown in Fig.~\ref{critical}, where the corresponding harmonic values are displayed for comparison. At 0K, including zero-point motion results in $\sigma_P^q\approx650$ MPa, which we use as a common point (joined by dashed lines) for all the curves.
The critical stress vanishes at $T=700$K, which represents the athermal limit within our model. This is higher than experimentally observed but our results show that the mere consideration of temperature effects leads to important changes in the dynamic picture of screw dislocations. As an example, at 100K, $\sigma^*\approx400$ MPa, which represents a 60\% reduction with respect to the value of $\sigma_P$ at 0K.

The data provided in Fig.\ \ref{critical} are a central result of this paper and reveal two main behaviors related to the Peierls potential. As referred to earlier, tensile tests in pure Fe show that the temperature dependence of the flow stress is characterized by the critical stress at (very) low temperatures ($<4$ K) and the athermal limit at high temperatures. In the absence of quantum corrections, atomistic calculations, even of the most accurate kind, fail to predict the lower temperature limit, while there is no numerical work available providing information for the higher one.
The present calculations show that a formal treatment of the free energy may account for dramatic reductions in both limits using conventional interatomic force fields. Indeed, these calculations provide a closed set of data for defining a temperature-dependent substrate potential to be used in higher level models such as line tension, line-on-substrate, kinetic Monte Carlo, etc. We believe the implications of this work to be of importance to all bcc metals.
It is worth emphasizing that the non-sinusoidal nature of $U_P$ from the force field employed here is not a weakening aspect of this work because finite-temperature trajectories sample paths in phase space that are not affected by the existence of the split core configuration at 0K. This may explain why most molecular dynamics simulations of screw dislocation motion using the present interatomic potential show only \emph{correlated} (in the sense of Gordon {\it et al.}~\cite{gordon2010}) formation of kink pairs \cite{domain2005,gilbert2011} at finite temperatures.


To conclude, we have presented a free energy map of the $\small{\frac{1}{2}\langle111\rangle}$ screw dislocation core transition on $\{110\}$ planes. The calculations have been done using constrained reaction coordinate dynamics and reveal a drastic reduction in free energy barrier and critical stress with increasing temperature. Our results can serve as yet another platform from which to interpret the discrepancies observed between atomistic simulations and macroscopic flow stress measurements.

The authors gratefully acknowledge helpful discussions with D. Rodney, S. Dudarev, and P. Derlet.
This work was performed under the auspices of the U.S. Department of Energy by Lawrence Livermore National Laboratory under Contract DE-AC52-07NA27344.
This work was part-funded by the RCUK Energy Programme under grant EP/I501045 and the European Communities under the contract of Association between EURATOM and CCFE. The views and opinions expressed herein do not necessarily reflect those of the European Commission.

\end{document}